\documentclass[useAMS,usenatbib]{mn2e}

\usepackage{threeparttable}
\usepackage{pdflscape}
\usepackage{graphicx}
\usepackage{longtable}
\usepackage{url}

\title[H$_{2}$O maser astrometry: AFGL 5142]{Trigonometric distance and proper motions of H$_2$O maser bowshocks in AFGL 5142}
\author[R. A. Burns]{R. A. Burns$^{1,2}$\thanks{E-mail:
burns@jive.eu}, T. Handa$^{2}$, H. Imai $^{2}$, T. Nagayama $^{3}$, and T. Omodaka$^{2}$ \newauthor  T. Hirota$^{4,5}$, K. Motogi$^{6}$, H. J. van Langevelde$^{1,7}$, W. A. Baan,$^{2,8}$\\
$^{1}$Joint Institute for VLBI ERIC (JIVE), Postbus 2, 7990 AA Dwingeloo, the Netherlands\\
$^{2}$Graduate School of Science and Engineering, Kagoshima University, 1-21-35 K\^orimoto, Kagoshima 890-0065, Japan\\
$^{3}$Mizusawa VLBI Observatory, National Astronomical Observatory of Japan,
2-12 Hoshigaoka-cho, Mizusawa-ku, Oshu, Iwate, Japan 023-0861\\
$^{4}$Mizusawa VLBI Observatory, National Astronomical Observatory of Japan,
Osawa 2-21-1, Mitaka, Tokyo 181-8588, Japan\\
$^{5}$Department of Astronomical Sciences, SOKENDAI, Osawa 2-21-1, Mitaka, Tokyo 181-8588, Japan\\
$^{6}$Graduate School of Sciences and Technology for Innovation,
Yamaguchi University.Yoshida 1677-1, Yamaguchi 753-8512, Japan\\
$^{7}$Sterrewacht Leiden, Leiden University, Postbus 9513, 2300 RA Leiden, the Netherlands\\
$^{8}$ASTRON, PO Box 2, NL-7990 AA Dwingeloo, the Netherlands}
\begin{document}

\date{Accepted 1988 December 15. Received 1988 December 14; in original form 1988 October 11}

\pagerange{\pageref{firstpage}--\pageref{lastpage}} \pubyear{2002}

\maketitle

\label{firstpage}

\begin{abstract}

\noindent We present the results of multi-epoch VLBI observations of water masers in the AGFL 5142 massive star forming region. We measure an annual parallax of $\pi=0.467 \pm 0.010$ mas, corresponding to a source distance of $D=2.14^{+0.051}_{-0.049}$ kpc. 
Proper motion and line of sight velocities reveal the 3D kinematics of masers in this region, most of which associate with millimeter sources from the literature.
In particular we find remarkable bipolar bowshocks expanding from the most massive member, AFGL 5142 MM1, which are used to investigate the physical properties of its protostellar jet. 
We attempt to link the known outflows in this region to possible progenitors by considering a precessing jet scenario and we discuss the episodic nature of ejections in AFGL 5142.

\end{abstract}



\section{Introduction}

Massive star formation is a challenging field of observational astronomy, with massive young stellar objects (MYSOs) residing in deeply embedded environments, often at great distances from Earth.
From a theoretical perspective, high- and low-mass star formation were once considered distinctly different, with MYSOs initiating nuclear fusion while still deeply embedded in their parent cores, accretion persisting into the main sequence, in addition to producing strong radiation capable of limiting accretion onto the central star (see reviews by \citealt{Zinnecker07,Tan14}).

However, with recent advances in instrument sensitivity, resolution and data reduction techniques used in astronomy there has emerged a gradual convergence in the observed features present in low- and high-mass star formation. 
MYSOs are increasingly found to harbour disk-jet systems - a feature ubiquitous to low mass star formation. Phenomena such as circumstellar disks \citep{Beltran04,Hirota14,Chen16,Ilee16}, episodic ejection \citep{Burns16b} and even jet rotation \citep{Burns15a} are now becoming common targets of investigation in MYSOs.

In disk-jet systems, ejection rates correlate with accretion rates \citep{Corc98,Garatti15} - thus the ejection history of an MYSO allows inference of its accretion history - which is itself unobservable in a practical sense due to the timescales involved. Accretion history, based on ejection history, can therefore be used to compare low- and high-mass star formation.

In addition to accretion mechanisms, outflows serve as another point of comparison - with the outflows of massive stars considered to be less collimated than those of low mass stars \citep{Wu04}. 
However, such conclusions are often based on surveys of large scale outflows which could have formed via entrainment by a collimated, precessing jet as was observed in IRAS 20126+4104 \citep{Shep00}. Evidently, all physical scales must be considered when comparing the outflows of low- and high-mass YSOs.

AFGL 5142 is a massive star forming region which has multiple outflows on multiple scales.
The region contains nine millimeter cores, by far the most predominant of which being MM1 and MM2 which exhibit hot core chemistry \citep{Zhang07,Palau11,Palau13}. MM1 has a mass of about 6.5 M$_{\odot}$ \citep{Liu16} and houses an embedded massive star, indicated by the presence of 6.7 GHz methanol masers which trace an infalling disk \citep{Goddi11}. An ionized bipolar jet extends to the NW-SE - at an angle near perpendicular to the disk. Water masers tracing the leading tip of the ionised jet reveal expanding motions in the line of sight and sky plane \citep{Goddi06,Goddi11}, thus giving AFGL 5142 MM1 the image of a prototypical disk-jet system in an MYSO. Furthermore, the combination of features associated with both low- and high-mass star formation make AFGL 5142 MM1 a good target for a comparative investigation.

About 1$^{\prime \prime}$ to the south is MM2, with a mass of about 6.2 M$_{\odot}$ \citep{Liu16} but no centimeter emission or 6.7 GHz methanol masers. Both MM1 and MM2 exhibit similar hot molecular core chemical compositions \citep{Zhang07,Palau11}, suggesting similar evolutionary stages. Water masers have also been detected near AFGL 5142 MM2 \citep{Hunter95,Goddi06}.

Interferometric observations of the AFGL 5142 region in CO ($2-1$) \citep{Zhang07,Palau11}, SiO ($2-1$) and HCO$^+$ ($1-0$) \citep{Hunter99} reveal at least four distinct collimated molecular outflows on $10^{ \prime \prime}$ scales; outflows A, B, C \citep{Zhang07} and outflow D \citep{Palau11}. Several of these outflows intersect both MM1 and MM2, barring a simple assignment of progenitors. Single dish observations reveal the presence of larger, arc-minute scale outflows traced in CO ($2-1$) which also intersect the MM1 and MM2 region \citep{Hunter95} these propagate primarily in the skyplane. Recently \citet{Liu16} also find a $\sim 9^{ \prime \prime}$ length, extremely wide-angle bipolar outflow (EWBO), with an opening angle of $\sim$180$^{\circ}$, driven from the MM1 - MM2 region, leading the authors to interpret its formation as being driven by a precessing jet rather than entrainment at wide angles from a narrow jet.


In this work we investigate the outflows seen in AFGL 5142 at the smallest scales by performing new multi-epoch very long baseline interferometry (VLBI) observations of water masers at 22 GHz. VLBI is an ideal observational approach to investigating protostellar ejections as multi-epoch observations provide sky-plane and line of sight information, thus revealing the 3D kinematics of gas in the vicinity of MYSOs.
Our investigation aims to provide observational characterisation of outflow behaviour (launching mechanisms and episodic ejection) for a prototypical MYSO, to provide comparison between low- and high-mass formation mechanisms.
Our report concentrates on the kinematics and energetics of the jets associated with AFLG 5142 MM1. We use our findings to offer a reinterpretation of the recent history of jets and outflows in this region and attempt to assign progenitors to each of the known outflows from the literature. Via annual parallax we also provide the first precise measurement of the distance to AFGL 5142; a quantity essential for converting observations to physical quantities such as ejection velocities, ages and momentum rates.

\section{Observations and data reduction}

VLBI observations of AFGL 5142 were carried out with VERA (VLBI exploration of radio astrometry). All observations were conducted in dual-beam mode \citep{dual} with beams centered on the maser target, AFGL 5142, and a reference quasar, J0533+3451. For both beams, left-hand circular polarisation signals were recorded at each of the 4 VERA station. Data were collected and correlated using the Mitaka FX correlator \citep{Chikada91}, adopting a rest frequency of 22.235080 GHz. For AFGL 5142 data we used a phase tracking center of
$(\alpha, \delta)_{\mathrm{J}2000.0}=(05^{\mathrm{h}}30^{\mathrm{m}}48^{\mathrm{s}}.01733742$,
+33$^{\circ}$47'54".56750)
and for J0533+3451 data we set a phase tracking center at 
$(\alpha, \delta)_{\mathrm{J}2000.0}=(05^{\mathrm{h}}33^{\mathrm{m}}12^{\mathrm{s}}.76510600$,
+34$^{\circ}$51'30".336995). Further phase corrections, including more accurate atmospheric models and antennae positions than those used at correlation, were made and applied post-calibration.

\begin{table}
\scriptsize
\caption{\small Summary of observations\label{obs}}
\begin{center}
\begin{tabular}{cccc}
\hline
Epoch& Observation & Modified& Number of \\
number& date& Julian date &features\\ \hline
1&  21 Apr 2014 & 56768 & 12\\
2&  20 May 2014 & 56797 & 9\\
3& 2 Oct 2014 &   56932 & 17\\ 
4& 25 Nov 2014 &  56986 & 22\\
5& 31 Jan 2015 &  57053 & 24\\
6& 29 Mar 2015 &  57110 & 25\\
7& 29 May 2015 &  57171 & 19\\
\hline
\end{tabular}
\end{center}
\end{table}

\begin{figure}
\begin{center}
\includegraphics[scale=0.88]{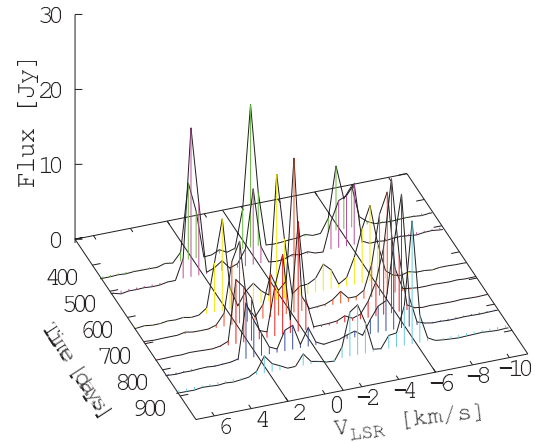}
\caption{\small Scalar averaged cross-power spectra of the maser emission in AFGL 5142 as a function of time, coloured arbitrarily. 
\label{spectra}}
\end{center}
\end{figure}

The total correlator bandwidth of 240 MHz was shared into 16 intermediate frequencies (IFs). One IF was allocated to the maser data with a bandwidth of 8 MHz and 15.63 kHz channel spacing, providing a 0.12 km s$^{-1}$ velocity spacing. The remaining 15 IFs were allocated to the data of the quasar reference source - one IF had similar properties to the maser data while the other 14 IFs had 16 MHz bandwidth and spacings of 125 kHz. The 15 quasar data IFs were manipulated into common form and merged - resulting in almost continuous frequency coverage spanning 232 MHz.

Data reduction was performed using AIPS (Astronomical Image Processing System), developed by National Radio Astronomy Observatory. VLBI phase referencing data reduction made use of the \emph{inverse phase referencing} technique, customised for the dual-beam data of VERA. The technique was first introduced in \citet{Imai12} and further development, in addition to a guide to its implementation, is given in \citet{Burns15a}.

\begin{table*}
\vspace{-0.2cm}
\begin{center}
\caption{\small The general properties of H$_{2}$O masers in AFGL 5142, detected with VERA. \label{TAB}}
\begin{tabular}{cclcccccccc}
\hline
Maser&$V_{\rm LSR}$&Detected &$\Delta \alpha \cos \delta$&$\Delta \delta$ &$\mu_{\alpha}\cos\delta$&$\mu_{\delta}$& F$_{\rm int}$&$\pi$\\ 
ID &(km s$^{-1}$)&epochs&(mas)&(mas)&(mas yr$^{-1}$)&(mas yr$^{-1}$)&(Jy)& (mas)\\
\hline
N.W. 1	&	$	2.47	$	&	1234567 &	-35.216	&	35.133	&	$	-0.55	\pm	0.03	$	&	$	0.89	\pm	0.26	$	&23.67&	$	0.460 \pm 0.014	$	\\
N.W. 2	&	$	0.42	$	&	1234567 &	-39.637	&	29.447	&	$	-1.03	\pm	0.02	$	&	$	0.23	\pm	0.11	$	&11.09&	$	0.479 \pm 0.021	$	\\
N.W. 3	&	$	3.49	$	&	****567 &	-28.365	&	38.437	&	$	0.07	\pm	0.27	$	&	$	0.16	\pm	0.05	$	&1.27&	$		$	\\
N.W. 4	&	$	3.07	$	&	1*34*67 &	-37.371	&	32.979	&	$	-0.58	\pm	0.08	$	&	$	1.07	\pm	0.08	$	&2.99&	$		$	\\
N.W. 5	&	$	1.81	$	&	**34567 &	-38.121	&	31.934	&	$	-0.19	\pm	0.48	$	&	$	1.51	\pm	0.80	$	&4.13&	$		$	\\
N.W. 6	&	$	1.24	$	&	****567 &	-38.838	&	30.803	&	$	-1.28	\pm	0.06	$	&	$	-0.21	\pm	0.11	$	&4.14&	$		$	\\
N.W. 7	&	$	0.86	$	&	**3*567 &	-41.625	&	24.954	&	$	-0.77	\pm	0.12	$	&	$	0.87	\pm	0.31	$	&2.22&	$		$	\\
N.W. 8	&	$	-0.02	$	&	****567 &	-45.198	&	19.198	&	$	-0.73	\pm	0.42	$	&	$	1.98	\pm	0.21	$	&2.46&	$		$	\\
\hline
S.E. 1	&	$	-6.69	$	&	123456* &	162.415	&	-232.163	&	$	1.13	\pm	0.14	$	&	$	-1.59	\pm	0.04	$	&4.83&	$		$	\\
S.E. 2	&	$	-6.20	$	&	1234567	&	156.283	&	-237.916	&	$	1.93	\pm	0.03	$	&	$	-1.16	\pm	0.03	$	&12.87&	$	0.479 \pm 0.022	$	\\
S.E. 3	&	$	-6.71	$	&	1**4567	&	147.805	&	-222.823	&	$	1.38	\pm	0.19	$	&	$	-0.83	\pm	0.11	$	&1.77&	$		$	\\
S.E. 4	&	$	-4.93	$	&	**34567	&	134.663	&	-228.13 	&	$	0.84	\pm	0.04	$	&	$	-1.19	\pm	0.08	$	&5.31&	$		$	\\
S.E. 5	&	$	-5.36	$	&	***4567 &	117.507	&	-230.753	&	$	0.36	\pm	0.04	$	&	$	-1.08	\pm	0.02	$	&2.72&	$		$	\\
S.E. 6	&	$	-6.91	$	&	****567	&	160.06	&	-235.217	&	$	1.97	\pm	0.70	$	&	$	-1.67	\pm	0.26	$	&2.52&	$		$	\\
S.E. 7	&	$	-4.08	$	&	**3****	&	121.616	&	-227.994	&	$	-		$	&	$	-		$	&1.41&	$		$	\\
S.E. 8	&	$	-7.05	$	&	*****6*	&	131.305	&	-229.803	&	$	-		$	&	$	-		$	&5.80&	$		$	\\
S.E. 9	&	$	-7.05	$	&	*****6*	&	159.065	&	-236.213	&	$	-		$	&	$	-		$	&0.88&	$		$	\\
S.E. 10	&	$	-5.77	$	&	**3****	&	54.856	&	-143.514	&	$	-		$	&	$	-		$	&2.15&	$		$	\\
\hline
F.S. 1	&	$	-5.78	$	&	1234567	&	11.783	&	-1241.213	&	$	1.42	\pm	0.03	$	&	$	-0.43	\pm	0.07	$	&13.98&	$	0.450 \pm 0.029	$	\\
F.S. 2	&	$	-4.41	$	&	***4567	&	12.405	&	-1227.991	&	$	1.64	\pm	0.08	$	&	$	0.05	\pm	0.21	$	&17.80&	$		$	\\
F.S. 3	&	$	-5.15	$	&	123456*	&	13.319	&	-1236.244	&	$	1.71	\pm	0.04	$	&	$	0.71	\pm	0.18	$	&5.74&	$		$	\\
F.S. 4	&	$	-4.43	$	&	**34567	&	13.227	&	-1230.149	&	$	1.35	\pm	0.08	$	&	$	1.00	\pm	0.11	$	&4.67&	$		$	\\
F.S. 5	&	$	-5.54	$	&	1234567	&	11.416	&	-1225.99	&	$	1.13	\pm	0.09	$	&	$	1.44	\pm	0.23	$	&11.02&	$		$	\\
\hline
B	&	$	0.77	$	&	1*3456*	&	639.914		&	148.019	&	$	1.89	\pm	0.06	$	&	$	-0.61	\pm	0.04	$	&3.06&	$		$	\\
F.S.W.&	$	-3.04	$	&	**3456*	&	-3457.993	&	-1553.682&	$	-2.56	\pm	-0.46	$	&	$	0.01	\pm	0.18	$	&3.29&	$		$	\\
A	&	$	-3.18	$	&	123456*	&	6.68	&	-40.77	&	$	-		$	&	$	-		$	&1.27&	$		$	\\
\hline
&&&&&&& Simultaneous fit & $0.467 \pm 0.010$ \\
\hline
\end{tabular}
\begin{tablenotes}
\item{\footnotesize{Column (2)}: Line of sight velocities, with respect to the local standard of rest, are quoted as those measured for the first detection.}
\item{\footnotesize{Column (3)}: Numbers indicate detection in the corresponding epoch, while asterisk represents non-detection.}
\item{\footnotesize{Column (8)}: Integrated fluxes are given as the average of measurements from all epochs.}
\end{tablenotes}
\end{center}
\end{table*}

In short, data reduction involves solving frequency dependant phase terms (group delay) using bright calibrators in the wider 15 IF data set, while time dependant phase terms are solved using the narrow but bright line emission of the maser. Both sets of solutions are then applied to the reference source, J0533+3451, inversely giving the angular separation of the maser from the phase center. Assuming that the reference source is fixed at its ICRF coordinates we then obtain the astrometric position of the maser from the derived separations. Since phase solutions determined using the maser are applied to itself we also produce high quality, astrometrically accurate maser maps. Typically, maser maps achieved rms noise values of $100 - 200$ mJy beam$^{-1}$ while the wider band continuum data achieved typical rms values of $\sim 1$ mJy.

Maser maps were produced using the AIPS task CLEAN in automatic mode (DOTV = 0), with a synthesised beam of dimensions $1.0\times1.2$ mas. Masers were identified from the maps using an automated \emph{SAD} routine in AIPS, employing a detection signal to noise cutoff of 7. Following common nomenclature, a `spot' refers to an individual maser emission peak, imaged in a single spectral channel - while a maser `feature' refers to a collection of spots considered to emanate from the same physical maser cloud. Maser spots were grouped into a feature if they shared a spectral feature and collocated within a radius of 1 mas. The associations of spots to features, and the flux weighted astrometric positions of maser features, were determined using a basic FORTRAN code which also removes the offset introduced by the difference in coordinates between the phase tracking center and the reference maser used in each epoch, in addition to shifting the maser maps to the absolute reference frame.

\begin{figure*}
\begin{center}
\includegraphics[scale=1.42]{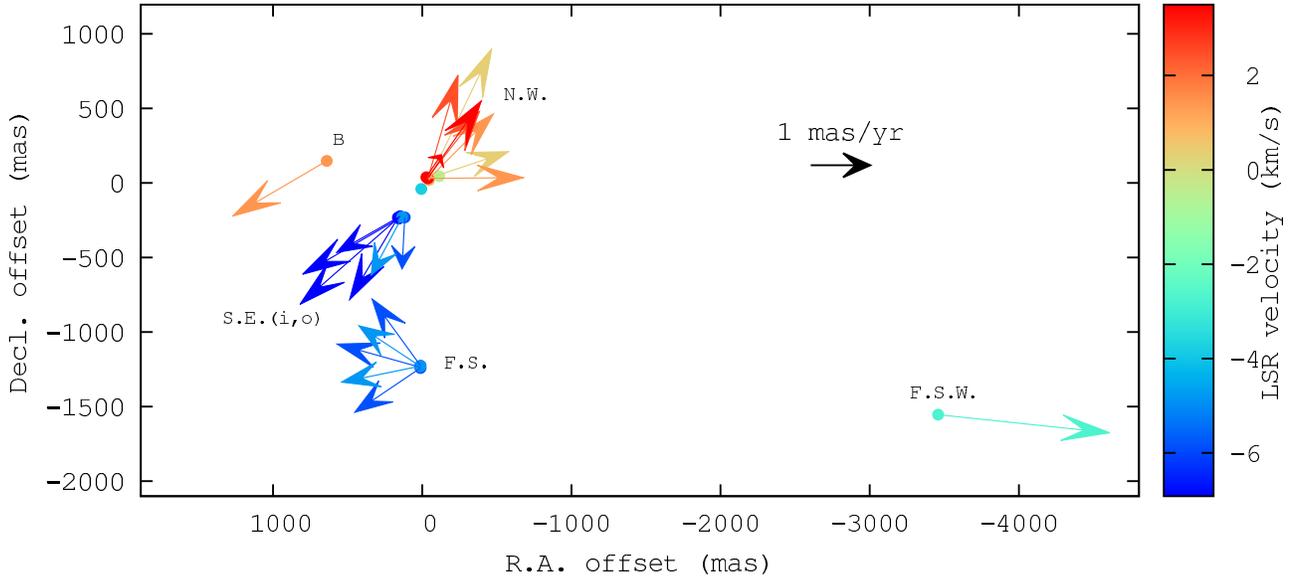}
\caption{\small Internal motions of water masers in AFGL 5142. Vector magnitudes indicate proper motion while colours indicate LSR velocity. Annotations are the same as those in Table~\ref{TAB}, with S.E. i and o referring to the `inner' and `outer' South West groups, respectively. The map origin is at the phase tracing center of the maser data.
\label{FULL}}
\end{center}
\end{figure*}

\null

\section{Results}

\subsection{Maser distribution and l.o.s. velocities}
A total of 27 independent maser features were detected with VERA during the observing calendar. Their coordinates, line of sight (l.o.s) velocities, detection frequencies and other properties are recored in Table~\ref{TAB}.
With regards to temporal variability, some maser features persisted through all epochs, while others exhibited more sporadic behaviour. No periodic behaviour was noted. Figure~\ref{spectra} shows the temporal changes in the overall emission spectrum, and the variability of individual maser features can be read from the detection frequencies in column 3 of Table~\ref{TAB}.  The distributions of all maser features detected by VERA are shown in Figure~\ref{FULL}.

\begin{figure}
\begin{center}
\includegraphics[scale=0.88]{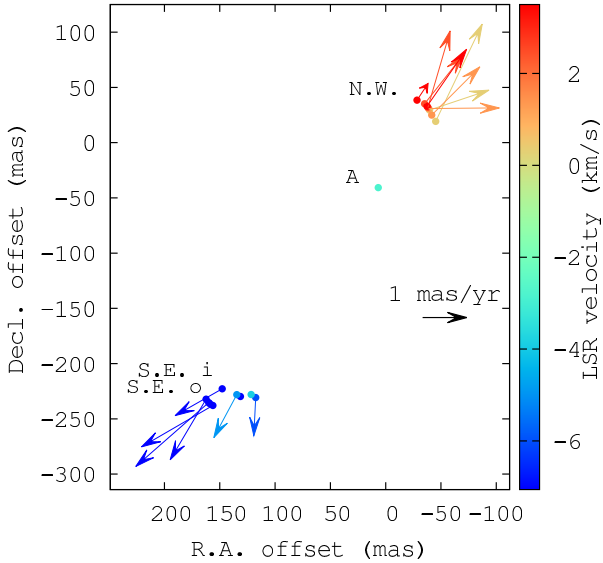}
\caption{\small Same as Figure~\ref{FULL} for the MM1 core region. 
\label{MM1}}
\end{center}
\end{figure}

Figure~\ref{MM1} shows a magnified view of the MM1 region. Masers trace three arc structures contained in a region of about $300 \times 300$ mas. There exists one maser arc to the North West of the system center, and two proximate and similarly oriented arcs lying to the South East, one closer and one further from the system center. We refer to these as the `N.W.', the `S.E. inner', and `S.E. outer' maser arcs, respectively - and are hence categorised accordingly in column 1 of Table~\ref{TAB}. The N.W. masers arcs are redshifted- and the S.E. masers blueshifted with respect to the l.o.s velocity of the region ($-1.1$ km s$^{-1}$ \citealt{Zhang07}).


Maser feature A (see Table~\ref{TAB}) is also associated with MM1. This maser does not associate with the aforementioned arcs, and rather, stands in the midway - near the system center, at map position $(\alpha, \delta) = (7,-41)$ mas. This maser exhibits some remarkable properties which will be discussed in a later publication.

Our observations did not detect all of the masers reported by \citet{Goddi06}. In their VLBA observations they find more water masers associated with MM1 which are located further to the N.W. We include their results in our investigation of the maser distribution in AFGL 5142 MM1 (see Section~\ref{epi}).


Water masers associated with MM2 are arranged in a single arc (see Figure~\ref{FSBOW}). Due to their location `far south' of the phase tracking center we refer to these masers as the F.S. group in Table~\ref{TAB}. These masers likely correspond to maser features `IIb' of \citet{Goddi06}. All F.S. masers have slightly blueshifted l.o.s. velocities.

We found one maser to the North East (`B' in Table~\ref{TAB}) of the phase tracking center - possibly associated with MM1 and probably corresponding to maser feature `IIr' of \citet{Goddi06}. One maser feature was found in the far South West (`F.S.W.'), associated with AFGL 5142 MM6.

\subsection{Fitting parallax and proper motion}

The astrometry of water masers observed with VERA was used to measure the annual parallax and resulting trigonometric distance of the AFGL 5142 region.
To assure the best possible precision for the distance estimate we performed parallax fitting only using maser features which were detected in all 7 epochs - spanning just over one year. Four maser features passed this criterion, two of which were in the N.W. bowshock, one in the S.E. bowshock and one in the F.S. bow shock. Parallax fitting was done simultaneously for the four maser features, assuming a common distance, to average and smooth out noise-like and structure induced errors. Additional error floors were added and reduced iteratively until a $\chi ^2$ value of unity was reached. Required error floors were 0.04 and 0.27 mas in R.A and Dec., respectively. \

\begin{figure}
\vspace{-0.666cm}
\begin{center}
\includegraphics[scale=0.85]{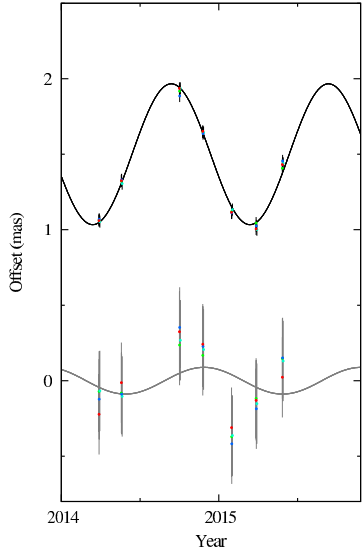}
\caption{\small Annual parallax motions of four maser features in AFGL 5142 in (\emph{above}) R.A. and (\emph{below}) Dec. directions. Maser features are coloured arbitrarily and error floors are represented by vertical bars. 
\label{PIE}}
\end{center}
\end{figure}

We estimate the annual parallax of AFGL 5124 to be $\pi=0.467 \pm 0.010$ mas, corresponding to a trigonometric distance of $D=2.14^{+0.051}_{-0.049}$ kpc, firmly placing it in the Perseus Arm. Fits are shown in Figure~\ref{PIE}. To confirm the reliability of our simultaneous fitting the parallaxes (\emph{see} Table~\ref{TAB}: column 7) and proper motions of each of the four maser features were then fit independently and were found to be consistent with each other and with the simultaneous fit.
Note that our consistent distances to MM1 and MM2 confirm that they are in close physical proximity.

Using Feature N.W. 1 as reference we obtained the relative proper motions of all other maser features that were detected in at least three epochs, of which there were 21. These relative motions were then converted to absolute proper motions (\emph{see} Table~\ref{TAB}: columns 6,7) using the absolute proper motion of Feature N.W. 1 from the astrometric fitting procedure described earlier. Proper motion errors were calculated as the quadrature sum of the standard deviation in the relative proper motion, and the proper motion uncertainty of Feature N.W. 1.

\subsection{Systemic motion and internal motions}

Absolute (observed) proper motions are a composition of the internal motions of masers in the frame of the driving source, and the apparent systemic motion of the source across the sky as governed by Galactic dynamics. In the case where a source's internal motions are expected to be symmetric the sum of all motion vectors should balance to zero - the systemic proper motion of the region is derived from the residual of this summation. In AFGL 5124 we used the proper motions of masers associated with the N.W. and S.E. maser bowshocks in MM1 (8 and 6 maser features, respectively) since the masers in this region are known to trace a symmetric bipolar ejection \citep{Goddi07}. The resulting systemic motion was ($\mu_{\alpha}\cos\delta$, $\mu_{\delta}$) = ($ +0.32\pm0.27$, $-0.22\pm0.47$) mas yr$^{-1}$, where errors were calculated as the quadrature sum of the average errors of all maser motions in each bowshock.

Subtracting the systemic proper motion from all absolute proper motions reveals the internal motions in the frame of the driving source, these are shown in Figures~\ref{FULL} and~\ref{MM1}.

Masers in MM1 trace arc shaped shocks which are expanding from the center of the system at an average velocity of 15 km s$^{-1}$. We interpret these as bow shocks tracing the protostellar jet reported by \citet{Goddi06,Goddi11} and discuss these in further detail later. The masers associated with MM2 also appear to trace an expanding arc. However the detection of only Easterly propagating components i.e. without a symmetric counterpart in the West, precludes detailed analysis of its kinematics.

Maser feature A could not be assigned a proper motion since its motion in the sky-plane was strongly non-linear.

\begin{figure}
\begin{center}
\includegraphics[scale=0.7]{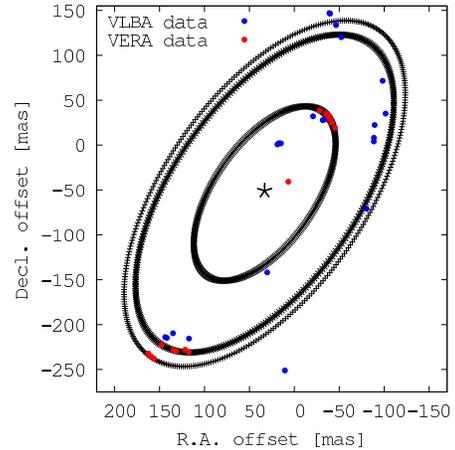}
\caption{\small Ellipses fit to the VERA data, assuming a common origin and indipendant ejections for N.W., S.E.i and S.E.o masers. VLBA data from 2004 \citep{Goddi06} are also shown.
\label{SINGLE}}
\end{center}
\end{figure}

\vspace{-0.1cm}

\subsection{Single vs Episodic ejection}
\label{epi}


To investigate the physical nature of ejections from MM1 we performed least squares fitting of elliptical jet shells to the maser distributions to evaluate their morphology and to determine the center position of the system. Since our maser data reveal two independent bowshocks in the S.E. lobe we must consider that ejections may occur in multiple episodes rather than a single ejection event, thus ellipses were fit under the condition that the three ejections, each traced by a bowshock, originate from a common center.
We find that the data are best fit to concentric ellipses with an origin at $(\alpha, \delta) = (33,-54)$ mas, shown as a star in Figure~\ref{SINGLE}. The center position coincides with the methanol disk reported by \citet{Goddi07} and is near to maser feature A which exhibits non-linear motions (\emph{see} Section~\ref{ALLVLBI}). The physical distances of the bowshocks from the system center were 240, 437 and 475 AU for the N.W., S.E. inner and S.E. outer masers, respectively, assuming that these bowshocks are close the plane of the sky. The shocked surfaces extend about $\sim 50$ AU in a direction perpendicular to the outflow direction, while their separations from the system center range from $240-475$ AU, indicating a high degree of collimation.

The difference in the ejection radii of the N.W. and S.E. bowshocks groups in our model can be interpreted either as separate asynchronus ejections expanding from the center at constant velocity, or as a quasi-single ejection event with non-equal propagation velocities - possibly due to a density gradient in the vicinity of the YSO. The former interpretation appears favoured considering the results of \citet{Goddi06} who find masers further in the N.W. direction than those observed with VERA (Figure~\ref{SINGLE}). These masers align well with the ellipses fit to the VERA data (eventhough the VLBA data were not used for fitting) and may thus be counterparts to the S.E. bow shocks.

Assuming expansion at constant velocity the ages of ejection events - i.e. the time it would have taken for ejecta to reach the positions of the observed bowshocks - can be estimated from their distance from the star divided by their outward motion. We measured average internal motions of 1.4 mas yr$^{-1}$ for each of the lobes, in opposite directions. For the ejections observed in AFGL 5142 MM1 we estimate ejection ages of 80, 145 and 159 years for the N.W., S.E. inner and S.E. outer bowshocks, respectively. In this scenario the S.E. bow shocks would be roughly twice the age of the N.W. bow shock, with 14 years interval between the ejections of the inner- and outer-S.E. ejections.


\section{Discussion}

\subsection{Physical properties of the jet}
\subsubsection{Outflow driving mechanism}
\label{ostriker}

Masers in AFGL 5142 MM1 trace arcs of shocked gas at the leading edge of a protostellar jet. 
Using the proper motions of masers measured with VLBI we can investigate the physical properties of the jet by comparison with published outflow models. VLBI maser data are well suited for comparison with the models of \citet{Lee01} and \citet{Ostriker01} who describe the motions of gas in the shocked region between a protostellar outflow and the ambient gas. In their works they compare two models; a jet driven bowshock and a stellar wind. 
Generally, the prevalence of each model can be determined by the amount of transverse motion observed near the head of the jet; in a jet driven outflow the transverse velocity is large as the jet material disperses at the bowshock while the transverse velocities in a stellar wind tend to zero at the tip since in a momentum driven stellar wind the gas propagates radially outward from the YSO, and thermal pressure is negligible.

\begin{figure*}
\begin{center}
\includegraphics[scale=1.42]{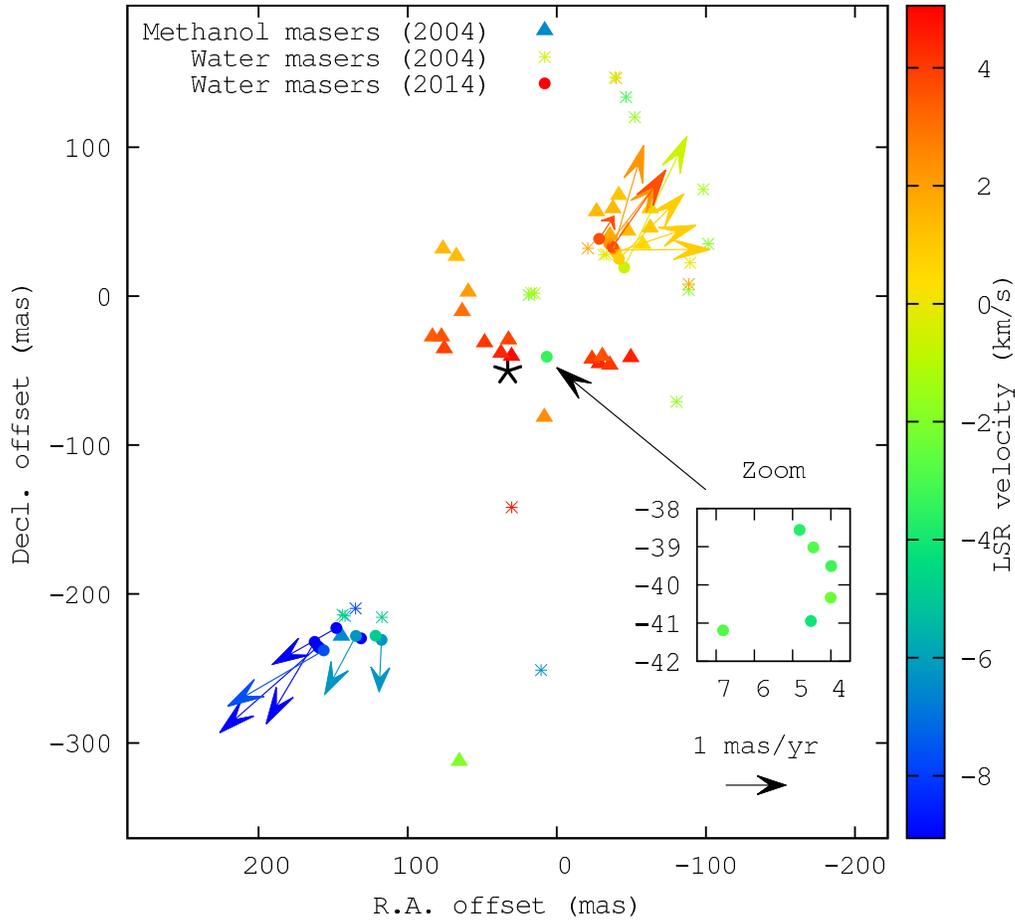}
\caption{\small Combined view of 22 GHz water masers (filled circles) observed with VERA in 2010 (\emph{this work}), 22 GHz water masers (asterisk) observed with the VLBA in 2004 \citep{Goddi06} and 6.7 GHz methanol masers (triangles) observed with the EVN in 2004 \citep{Goddi07}. The inset shows the trajectory of maser feature A, moving in a clockwise fashion. The trajectory of feature A and proper motions of other masers are all converted to the YSO frame. The black asterisk symbol indicates the approximate origin of the episodic ejections, estimated from least-squares fitting of ellipses to the VERA maser data.
\label{COMBINED}}
\end{center}
\end{figure*}

Using a jet position angle of $-35 ^{\circ}$ we measure transverse proper motions in a velocity range of $ 0 - 13$ km s$^{-1}$, with a mean value of $4$ km s$^{-1}$.
The large range of transverse motions indicate that the outflow in AFGL 5142 MM1 is dominated by jet-driven bowshock kinematics. Since masers in AFGL 5142 MM1 exclusively trace the leading surface of the outflow it is difficult to quantify any contribution from a stellar wind - whose influence is more easily seen along the body of the outflow.
Other similar works comparing VLBI determined maser proper motions with the models of \citet{Lee01} and \citet{Ostriker01} also find MYSO jets to be dominated by jet-driven bowshocks with some contribution from a lower velocity wind \citep{Sanna12,Burns16b}.


\subsubsection{Jet momentum rate}
\label{momrate}

In addition to investigating the driving mechanism, we can calculate the momentum rate (force) of the jet via the expression used by \citet{Goddi11}:

\vspace{-0.35cm}
\begin{eqnarray}
\dot{P}_{\rm outflow} = 1.5 \times 10^{-3}~ V^2_{10}~ R^2_{100}~ (\Omega / 4 \pi)~ n_8~ [{\rm M}_{\odot}~ {\rm km s}^{-1}~ {\rm yr}^{-1}].
\label{FORCE}
\end{eqnarray}

In which V$_{10}$ is the maser velocity in units of 10 km s$^{-1}$, R$_{100}$ is the radial distance of the masers from the driving source in units of 100 AU, $\Omega$ is the opening solid angle of the jet and $n_8$ is the volume density of ambient gas in units of $10^8$ cm$^{-3}$.
The following parameters used in the calculation were derived from our observations of the N.W. maser bowshock; $v=14$ km s$^{-1}$, $R = 240$ AU, and S.E. bowshock; $v=14$ km s$^{-1}$, $R = 475$ AU. Furthermore we take $\Omega = 1.1$ sr from the jet continuum observations of \citet{Goddi11}. 
Regarding the ambient density, $n_8$, we consider values in the range $10^7-10^9$ cm$^{-3}$ which are simulated initial ambient densities suitable for producing masers in shocks \citep{Kauf96}. We arrive at a range of momentum rates between $\dot{P} = 10^{-4} ~{\rm to}~ 10^{-2}$ M$_{\odot}$ km s$^{-1}$ yr$^{-1}$ for both bowshocks. These ranges are consistent with \citet{Goddi11} and the extended molecular outflows in AFGL 5142 MM1 (see Table~\ref{OFP}).

If we instead derive the jet opening angle directly from the maser data we arrive at $\Omega = 0.07$ sr and momentum rate ranges of $\dot{P} = 10^{-6} ~{\rm to}~ 10^{-4}$ M$_{\odot}$ km s$^{-1}$ yr$^{-1}$. The fundamental property that maser emission arises in regions of special physical conditions, and that they are observed preferentially along surfaces tangental to the observer, can complicate maser-derived estimates of the jet opening angle - except for cases where the jet head and body are more fully traced. As such, we argue that the jet opening angle derived from the continuum observations of \citet{Goddi11} is preferred.




\subsubsection{Primary jet velocity}

When the ram pressure of the jet is sufficient to balance that of the swept up material the relationship between the velocities of the primary jet, and that of the swept up gas, is given as $v_{j} \approx v_{ws} (1+ \sqrt{n / n_{j}}) $, adapted from \citet{Chernin94}. Where $v_{j}$ and $v_{ws}$ are the velocities of the primary jet and working surface (traced by the bowshock velocity), and $n$ and $n_j$ are the ambient and jet number densities, respectively. For the ambient density we use $n = 10^8$ cm$^{-1}$, adopted from shock models \citet{Hollenbach13}. Regarding the jet density, $n_j$, \citet{Garatti15} find primary jet electron densities of $n_e \sim 10^3$ to $10^4$ cm$^{-3}$ in their survey of MYSOs. Employing an ionisation fraction of $x_e = n_e/n_H = 0.1$ \citep{Hartigan94} the MYSO primary jet densities fall in the range of $n = 10^4$ to $10^5$ cm$^{-3}$. Finally, we estimate primary jet velocities in the range of $v_j \sim 100$ to $1000$ km s$^{-1}$. The jet velocity derived here for AFGL 5142 MM1 is consistent with other values for MYSOs from the literature \citep{Marti95,Curiel06,Sanna16,Guzman16,Kamenetzky16}.


\subsubsection{bowshock velocity gradient}

As a closing remark on the observationally determined jet physical properties in AFGL 5142 MM1, we note a shallow yet well structured velocity gradient across the N.W. bowshock masers (see Figure~\ref{VELGRAD}). This could be interpreted as a flattened and expanding ejection or as  rotation of the entrained material about the jet axis - a phenomenon widely sought in the context of massive star formation as it would suggest that outflows are driven by magneto-centrifugal jets (\emph{see also} \citealt{Burns15a}). Our data show that AFGL 5142 MM1 has potential as a target for future rotating jet investigations.

\subsubsection{Combined VERA, VLBA and EVN view of AFGL 5142 MM1}
\label{ALLVLBI}
AFGL 5142 has been observed by three different VLBI arrays; the VLBA \citep{Goddi06}, the EVN \citep{Goddi07}, and VERA (this work). These are presented together in Figure~\ref{COMBINED}.

The combined view of VLBI maser observations of AFGL 5142 depicts a prototypical disk-jet system; a structure ubiquitous in low mass star formation and becoming increasingly encountered in MYSOs with well studied cases including Orion Source I \citep{Hirota14}, IRAS 20126+4104 \citep{Moscadelli11,Chen16}. Water masers trace a narrow episodic jet which emanates at an angle near perpendicular to the protostellar disk traced by 6.7 GHz methanol masers - a transition exclusive to MYSOs.


Maser feature A, detected in this work only, exhibits a curved trajectory and lies close to the likely position of the star as was inferred independently by the center of kinematics of the 6.7 GHz methanol masers \citep{Goddi07}, the center of kinematics of the VLBA water masers \citep{Goddi06} and the concentric ellipses fit to VERA water masers - presented in Section~\ref{epi}. Non-linear motion implies the application of force - whether that be gravitationally or magnetically dominated we expect that the maser must be in close proximity to the central object.
The trajectory of feature A is shown in an inset in Figure~\ref{COMBINED} where motion progresses from the upper right of the inset in a clockwise direction. The trajectory has been shifted to the frame of the star.
We pursue a full decomposition of the trajectory of this maser in a future publication targeting the central $<100$ AU region.

Overall, the features highlighted by the combination of VLBA, EVN and VERA observations reveal a system comprising a circumstellar disk from within which a narrow and collimated jet is launched. These features resemble those representative of low mass YSOs. 
As we showed in Section~\ref{ostriker}, the outflow launching mechanism in AFGL 5142 MM1 also resembles those common to low-mass YSOs, while the outflow momentum rate of AFGL 5142 MM1 (Section~\ref{momrate}) is larger than those typically seen in low-mass stars \citep{Beuther02}. Episodic ejection, which in turn implies that accretion also occurs episodically, is another feature shared with low-mass disk-aided star formation. These points, taken together with the morphology and 3D kinematics of structures near the MYSO compiled from VLBI observations, reveals that AFGL 5142 MM1 depicts a `scaled up' picture of low-mass star formation.



\subsection{Molecular outflows in the AFGL 5142 region}
\label{ofs}

Previous attempts at locating the progenitors of the multiple outflows seen in AFGL 5142 struggled on account of there being more outflows (at least 4) than major millimeter cores (only two), and because most outflows intersect one or both of MM1 or MM2. 
We propose that several of the outflows in this region could be explained by episodic ejections from a slowly precessing outflow system. In a recent paper \citet{Liu16} consider that a precessing jet may be the best way to produce the extremely wide-angle bipolar outflow which forms the subject of their investigation. We share this view and extend on justifications described as follows.

\begin{figure}
\begin{center}
\includegraphics[scale=0.68]{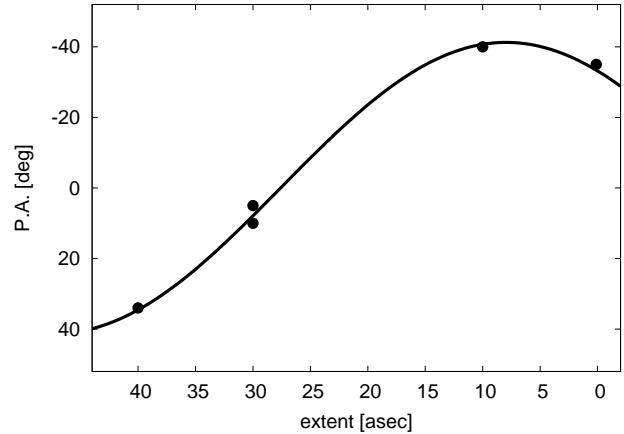}
\caption{\small The angular extent of outflows from Table~\ref{OFP} as a function of position angle. A sinusoid represents expected values for a slowly precessing outflow, fit to the data.
\label{PAVST}}
\end{center}
\end{figure}

Outflows A and C from \citet{Zhang07}, the `compact CO outflow' of \citet{Hunter95} and the SiO and HCO$^+$ outflows of \citet{Hunter99} share a trend whereby the redshifted lobe of the outflow extends to the North, while the blueshifted lobes extend to the south. We evaluate this trend in Figure~\ref{PAVST} which shows the P.A. of the aforementioned outflows as a function of angular extent, which we take as analogous to the outflow age (more extended outflows are older). Given that the age of the more extended outflows are a few $10^4$ years old \citep{Hunter95,Hunter99}, the precession has an approximate period of one revolution per $10^4 - 10^5$ years.
For this analysis we only use the Northern, redshifted outflows since the southernly directed blueshifted outflows often appear to truncate near MM2, which is known to coincide with the densest part of the molecular envelope (see Figure 4 of \citealt{Zhang07}). Outflow details used in the analysis and references to the original works are given in Table~\ref{OFP}.


\begin{figure*}
\begin{center}
\includegraphics[scale=1.99]{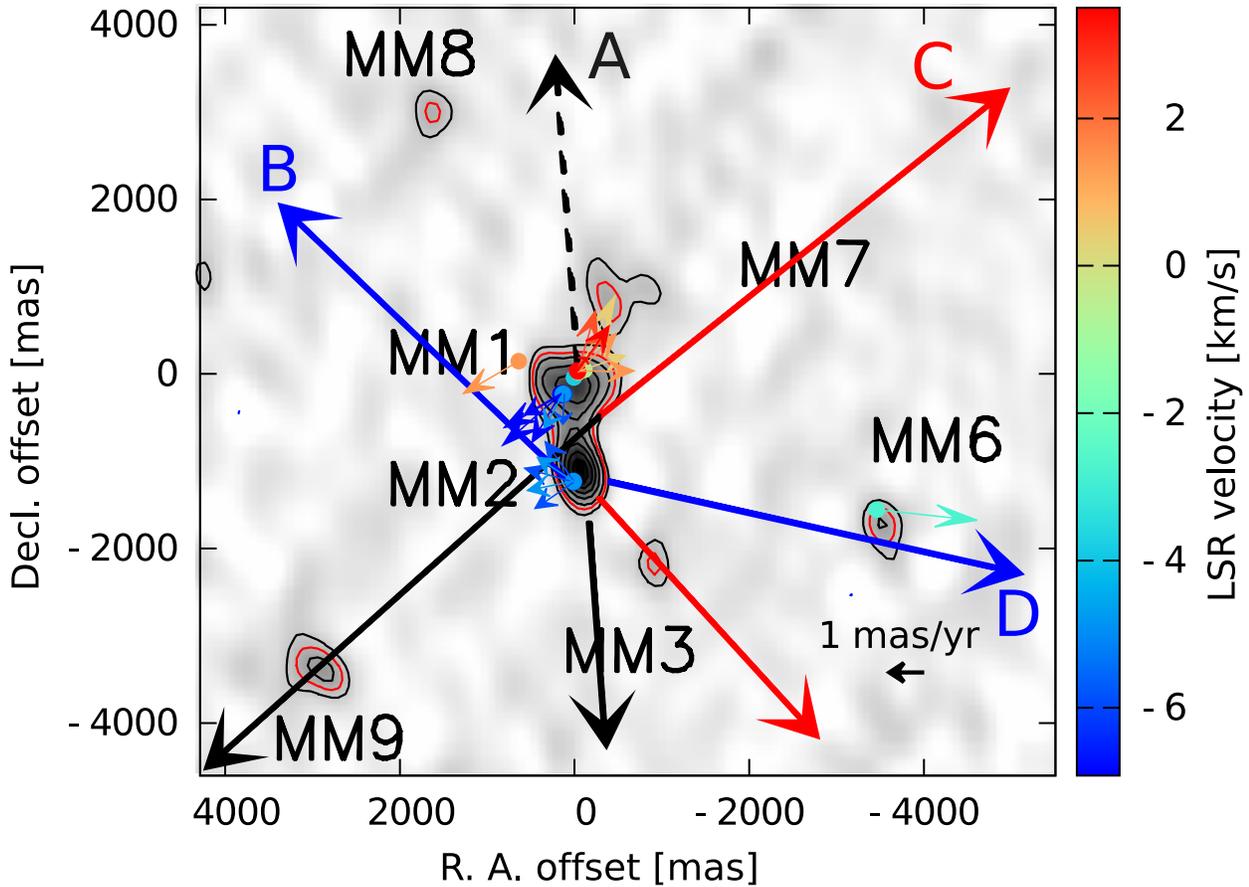}
\caption{\small Overlay showing water maser activity in the context of the millimeter core and molecular outflows. Small arrow vectors are the same as in Figure~\ref{FULL}. Greyscale shows the 1.3 mm continuum emission in the range of 0 - 0.07 mJy with contours at -4, 4, 8, 12, 16, 20 and 24 times the rms noise of 2.8 mJy beam $^{-1}$, from \citet{Palau13}, with the millimeter cores labeled. The red contour indicates the $6 \sigma$ detection adopted by \citet{Palau13}. Larger arrows with letters annotated indicate the directions of molecular outflows described in previous works \citep{Zhang07,Palau13}.
\label{SHOES}}
\end{center}
\end{figure*}

Figure~\ref{PAVST} demonstrates the plausibility that the outflows listed above emanated from a single, precessing jet system - which we can confidently attribute to MM1 thanks to works at the highest angular resolution (\citealt{Goddi06,Goddi11} and this work). 
To further test this hypothesis we consider the momentum rates of each outflow, which we calculate or re-evaluate from the literature using our trigonometric distance of 2.14 kpc and record in column 4 of Table~\ref{OFP}. All outflows under consideration have comparable momentum rates, supporting the hypothesis that they were driven by the same progenitor.



Regarding the remaining outflows, outflow D was likely driven by MM2 since its East-West orientation matches the cental axis of the bowshock in the F.S. masers (Figures~\ref{SHOES},\ref{FSBOW}), and also the proper motions of the F.S. and F.S.E masers (Figure~\ref{SHOES}). Note that both millimeter cores of MM1 and MM2 are elongated in the direction perpendicular to the assigned outflows, a known attribute of disk-jet systems \citep{Reid07}.

The velocity orientation and position angle of outflow B is opposite to those of outflows A and C, ruling out an association with MM1. Outflow B may be driven by MM3 which could be a low mass YSO \citep{Zhang07}. MM3 sits between the blue- and redshifted lobes of outflow B where \citet{Zhang07} detect weak 8.4 GHz continuum emission at the location of the millimeter core. The authors thus argue that MM3 may be of stellar nature, while the remaining millimeter cores, MM6, MM7, MM8 and MM9 can be explained as dust condensations swept up by molecular outflows, D, A/C, A and C, respectively.

\begin{table}
\vspace{-0.2cm}
\begin{center}
\scriptsize
\caption{\small Outflow parameters from this work and the literature. \label{OFP}}
\begin{tabular}{ccccc}
\hline
Tracer & $Length$ & P.A. & Momentum rate & Notes\\
       & $^{\prime \prime}$ & [$^{\circ}$] & [M$_{\odot}$ km s$^{-1}$ yr$^{-1}$] &\\
\hline
CO (2-1)     & 40 & 34 & $1.11 \times 10^{-4}$ & 1 \\
HCO$^+$ (1-0)& -  & -  & $4.52 \times 10^{-3}$ & 2 \\
SiO (2-1)    & 30 & 10 &                     - & 2 \\
CO (2-1)     & 30 & 5  & $2.85 \times 10^{-3}$ & `Outflow A', 3 \\
CO (2-1)     & 10  &-40 & $2.38 \times 10^{-3}$ & `Outflow C', 3, 4\\
H$_2$O $(6_{16}-5_{23})$&0.1&-35& $10^{-4} ~{\rm to}~ 10^{-2}$ & this work\\
\hline
\end{tabular}
\begin{tablenotes}
\item{Column (5)}: 1 - \citet{Hunter95}; 2 - \citet{Hunter99}; 3 - \citet{Zhang07}; 4 - \citet{Palau11}
\end{tablenotes}
\end{center}
\end{table}

\subsection{Episodic ejection in AFGL 5142}

Bowshocks trace the leading edges of new ejection events. As such, multiple bowshocks imply a history of multiple ejections. In Section~\ref{epi} we present evidence of episodic ejection operating on timescales of $10^1$ years, based on maser data. In Section~\ref{ofs} we argue that several of the molecular outflows in AFGL 5142 appear to stem from AFGL 5142 MM1.
Regarding the dynamical timescales of these molecular outflows, the older, more extended outflows from MM1 are a few $10^4$ years old \citep{Hunter95,Hunter99}, while there are outflows associated with MM1 on 4 different scales (Table~\ref{OFP}) - corresponding to ejection timescales of every few $10^4$ years.

Since outflow activity is expected to correlate with accretion activity \citep{Garatti15} the $10^4$ year timescale of ejections could indicate alternating periods of accretion bursts and quiescence in AFGL 5142 MM1, consistent with timescales predicted for episodic accretion in low- \citep{Stamatellos11} and high-mass stars \citep{Meyer17}. The short episodic ejection seen on a 10 year timescales (Section~\ref{epi}) would likely originate in the inner disk.


The concept of episodic accretion is only recently being considered in the framework of massive star formation, though its ability to deal with the "luminosity problem" and suppress the accretion inhibiting radiation field show great potential in explaining how massive stars accumulate their mass. AFGL 5142 presents a promising observational target for studies of episodic ejection and accretion in MYSOs.

\section{Conclusions}

Using VERA, a dual-beam VLBI array dedicated to astrometry, we measured an annual parallax for the AFGL 5142 massive star forming region of $\pi=0.467 \pm 0.010$ mas, corresponding to a trigonometric distance of D=$2.14^{+0.051}_{-0.049}$ kpc.

Using the latest VLBI H$_2$O maser data we analyse a recent ejection from the AFLG 5142 MM1 region which produced prototypical bi-polar bowshocks that expand from a common center in a circumstellar disk. We use the 3D kinematics of the maser bowshocks to investigate the physical properties of the protostellar jet.

One of the masers near the likely position of the star exhibits clear non-linear motion. The immediate region around the star requires further deep observations to provide vital context for this very unusual finding.

Combining our data with those of other published VLBI observations and through proper motion analysis we find that AFGL 5142 MM1 is forming as a disk-jet system with evidence of episodic ejection, giving it the appearance of `scaled up' low-mass star formation.


Finally, we reinterpret the outflow history of AFGL 5142 MM1 as formation via an episodic, precessing outflow. We were able to suggest progenitor allocations for all outflows in AFLG 5142 known from the current literature.

\section*{Acknowledgments}

R.B. would like to acknowledge the Ministry of Education, Culture, Sports, Science and Technology (MEXT) Japan for support as part of the Monbukagakusho scholarship.

We kindly thank the referee for providing helpful feedback and guidance during the review of this work.

\def\ref@jnl#1{{\rmfamily #1}}%
\newcommand\aj{\ref@jnl{AJ}}%
\newcommand\araa{\ref@jnl{ARA\&A}}%
\newcommand\apj{\ref@jnl{ApJ}}%
\newcommand\apjl{\ref@jnl{ApJ}}%
\newcommand\apjs{\ref@jnl{ApJS}}%
\newcommand\ao{\ref@jnl{Appl.~Opt.}}%
\newcommand\apss{\ref@jnl{Ap\&SS}}%
\newcommand\aap{\ref@jnl{A\&A}}%
\newcommand\aapr{\ref@jnl{A\&A~Rev.}}%
\newcommand\aaps{\ref@jnl{A\&AS}}%
\newcommand\azh{\ref@jnl{AZh}}%
\newcommand\baas{\ref@jnl{BAAS}}%
\newcommand\jrasc{\ref@jnl{JRASC}}%
\newcommand\memras{\ref@jnl{MmRAS}}%
\newcommand\mnras{\ref@jnl{MNRAS}}%
\newcommand\pra{\ref@jnl{Phys.~Rev.~A}}%
\newcommand\prb{\ref@jnl{Phys.~Rev.~B}}%
\newcommand\prc{\ref@jnl{Phys.~Rev.~C}}%
\newcommand\prd{\ref@jnl{Phys.~Rev.~D}}%
\newcommand\pre{\ref@jnl{Phys.~Rev.~E}}%
\newcommand\prl{\ref@jnl{Phys.~Rev.~Lett.}}%
\newcommand\pasp{\ref@jnl{PASP}}%
\newcommand\pasj{\ref@jnl{PASJ}}%
\newcommand\qjras{\ref@jnl{QJRAS}}%
\newcommand\skytel{\ref@jnl{S\&T}}%
\newcommand\solphys{\ref@jnl{Sol.~Phys.}}%
\newcommand\sovast{\ref@jnl{Soviet~Ast.}}%
\newcommand\ssr{\ref@jnl{Space~Sci.~Rev.}}%
\newcommand\zap{\ref@jnl{ZAp}}%
\newcommand\nat{\ref@jnl{Nature}}%
\newcommand\iaucirc{\ref@jnl{IAU~Circ.}}%
\newcommand\aplett{\ref@jnl{Astrophys.~Lett.}}%
\newcommand\apspr{\ref@jnl{Astrophys.~Space~Phys.~Res.}}%
\newcommand\bain{\ref@jnl{Bull.~Astron.~Inst.~Netherlands}}%
\newcommand\fcp{\ref@jnl{Fund.~Cosmic~Phys.}}%
\newcommand\gca{\ref@jnl{Geochim.~Cosmochim.~Acta}}%
\newcommand\grl{\ref@jnl{Geophys.~Res.~Lett.}}%
\newcommand\jcp{\ref@jnl{J.~Chem.~Phys.}}%
\newcommand\jgr{\ref@jnl{J.~Geophys.~Res.}}%
\newcommand\jqsrt{\ref@jnl{J.~Quant.~Spec.~Radiat.~Transf.}}%
\newcommand\memsai{\ref@jnl{Mem.~Soc.~Astron.~Italiana}}%
\newcommand\nphysa{\ref@jnl{Nucl.~Phys.~A}}%
\newcommand\physrep{\ref@jnl{Phys.~Rep.}}%
\newcommand\physscr{\ref@jnl{Phys.~Scr}}%
\newcommand\planss{\ref@jnl{Planet.~Space~Sci.}}%
\newcommand\procspie{\ref@jnl{Proc.~SPIE}}%

\bibliographystyle{mn2e}
\bibliography{./R2_I05274.bib}


\appendix
\section{VLBI images of masers}

\begin{figure}
\begin{center}
\includegraphics[scale=0.52]{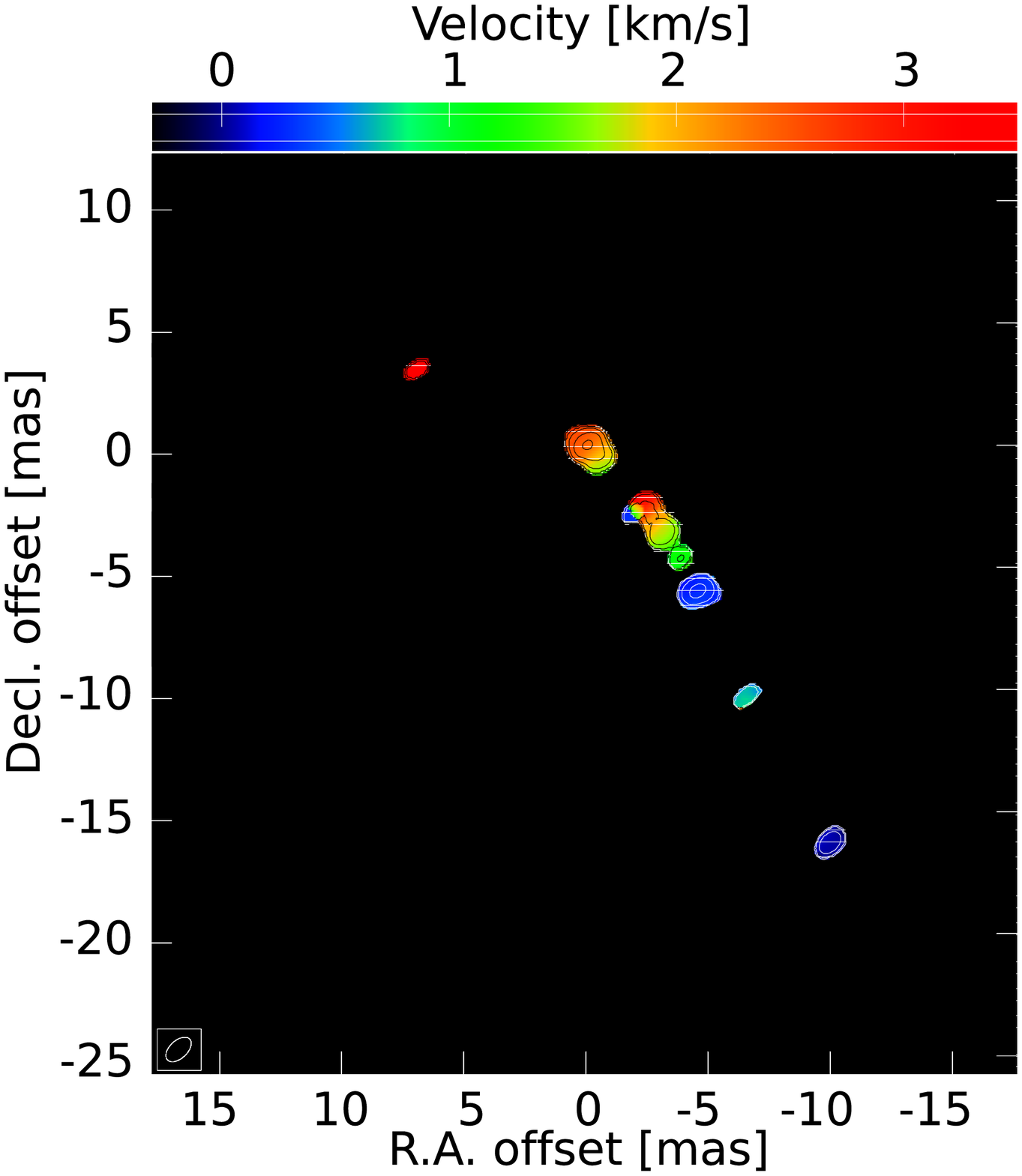}
\caption{\small Maser distributions and line of sight velocities in the N.W. bowshock.
\label{VELGRAD}}
\end{center}
\end{figure}

\begin{figure}
\begin{center}
\includegraphics[scale=0.3]{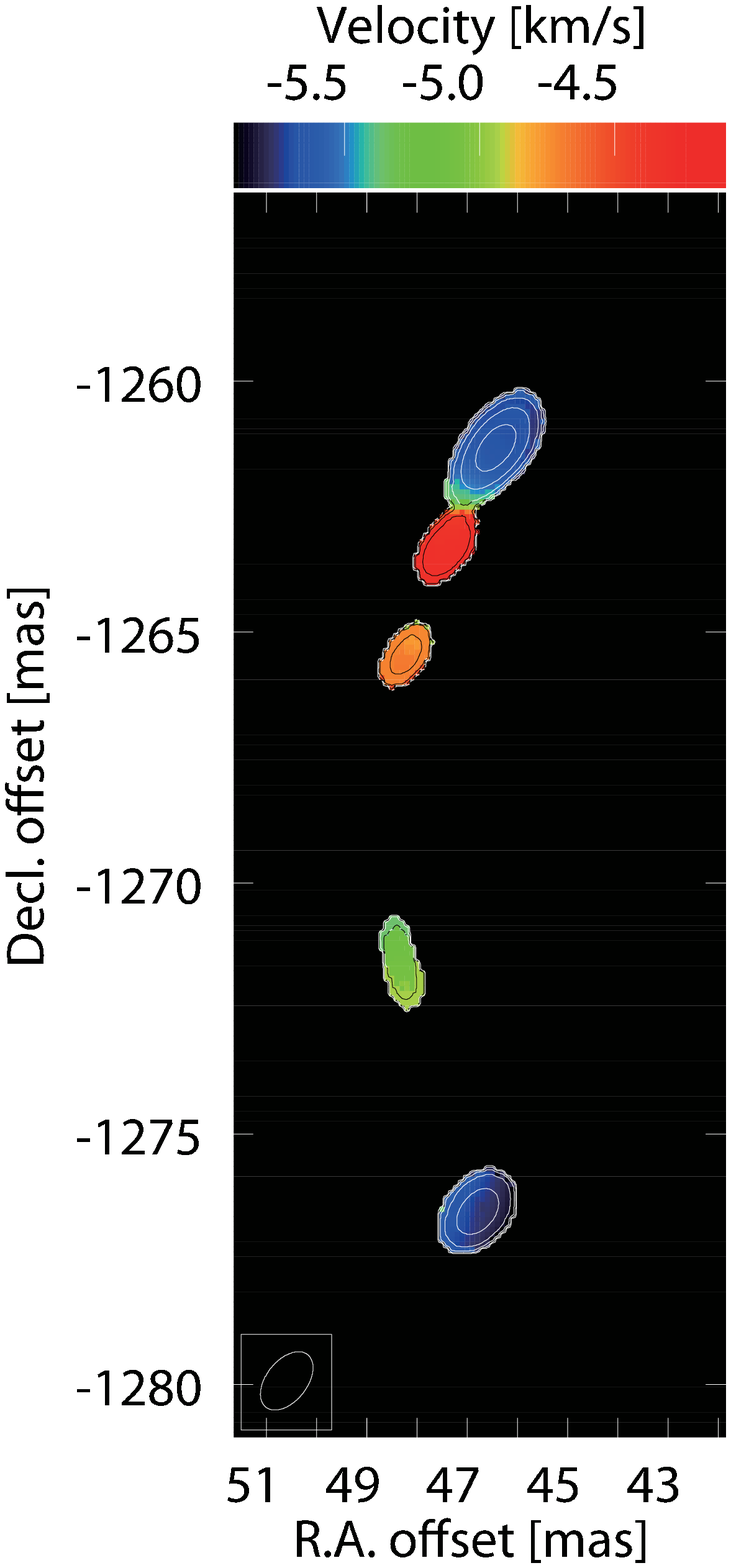}
\caption{\small Maser distributions and line of sight velocities in the F.S. bowshock.
\label{FSBOW}}
\end{center}
\end{figure}

\end{document}